\documentclass[preprint,eqsecnum,aps,nofootinbib]{revtex4}
\usepackage{amsfonts,amsmath,amssymb,amsthm}
\usepackage{latexsym}
\usepackage{bbm,bm}
\usepackage{graphicx}

\newcommand{\ket}[1]{\lvert #1 \rangle}
\newcommand{\bra}[1]{\langle #1 \lvert}
\newcommand{\beq}{\begin{equation}}
\newcommand{\eeq}{\end{equation}}
\newcommand{\beqs}{\begin{eqnarray}}
\newcommand{\eeqs}{\end{eqnarray}}

\begin{document}

\title{Entanglement Classification of Restricted Greenberger-Horne-Zeilinger Symmetric States in Four-Qubit System}

\author{DaeKil Park$^{1,2}$}

\affiliation{$^1$Department of Physics, Kyungnam University, Changwon
                  631-701, Korea  \\
             $^2$Department of Electronic Engineering, Kyungnam University, Changwon
                 631-701, Korea       
                      }

\begin{abstract}
Similar to the three-qubit Greenberger-Horne-Zeilinger (GHZ) symmetry we explore the four-qubit GHZ symmetry group and its subgroup
called restricted GHZ symmetry group. While the set of symmetric states under the whole group transformation is represented 
by three real parameters, the set of symmetric states under the subgroup transformation is represented by two real parameters. 
After comparing the symmetric states for whole and subgroup, the entanglement is examined for the latter set.
It is shown that the set has only two SLOCC classes, $L_{abc_2}$ and $G_{abcd}$. Extension to the multi-qubit system is briefly discussed.
\end{abstract}

\maketitle
\section{Introduction}

Quantum entanglement\cite{horodecki09} is the most important notion in quantum technology (QT) and quantum information theory (QIT).
As shown for last two decades it plays a crucial role in quantum teleportation\cite{teleportation},
superdense coding\cite{superdense}, quantum cloning\cite{clon}, and quantum cryptography\cite{cryptography}. It is also quantum entanglement, 
which makes the quantum computer outperform the classical one\cite{text,computer}. Thus, in order to develop QT and QIT it is essential to 
understand how to quantify and how to characterize the multipartite entanglement.

Since the quantum entanglement is a non-local property of given multipartite quantum state, it should be invariant 
under the local unitary (LU) transformations, i.e. the unitary operations acted independently on each of the subsystems.
If $\ket{\psi}$ and 
$\ket{\varphi}$ are in the same category in LU, one state can be obtained with certainty from the other one by means of local
operations assisted classical communication (LOCC)\cite{bennet00,vidal00}. This implies that $\ket{\psi}$ and $\ket{\varphi}$ can
be used, respectively, to implement the same task of QIT with equal probability of successful performance of the task. However, the 
classification of entanglement through LU generates infinite equivalence classes even in the simplest bipartite systems.


In order to escape this difficulty the classification through stochastic local operations and classical communication (SLOCC)
was suggested in Ref.\cite{bennet00}. If $\ket{\psi}$ and 
$\ket{\varphi}$ are in the same SLOCC class, one state can be converted into the other state with nonzero probability 
by means of LOCC. This fact implies that $\ket{\psi}$ and $\ket{\varphi}$ can
be used, respectively, to implement the same task of QIT although the probability of success for this task is different. 
Mathematically, if two $n$-party states $\ket{\psi}$ and $\ket{\varphi}$ are in the same SLOCC class, they are related to each other by 
$\ket{\psi} = A_1 \otimes A_2 \otimes \cdots \otimes A_n \ket{\varphi}$ with $\{A_j\}$ being arbitrary invertible local 
operators\footnote{For complete proof on the connection between SLOCC and local operations see Appendix A of Ref.\cite{dur00}.}. 
However, it is more useful to restrict ourselves to SLOCC transformation where all $\{A_j\}$ belong to 
SL($2$, $C$), the group of $2 \times 2$ complex matrices having determinant equal to $1$. 

The SLOCC classification was first examined in the three-qubit pure-state system\cite{dur00}. It was shown that the whole system consists of 
six inequivalent SLOCC classes, i.e., fully separable (S), three bi-separable (B), W, and Greenberger-Horne-Zeilinger (GHZ) classes.
Moreover, it is possible to know which class an arbitrary state $\ket{\psi}$ belongs 
by computing the residual entanglement\cite{ckw} and concurrences\cite{concurrence1} for its partially reduced states. Similarly, 
the entanglement of whole three-qubit mixed states consists of S, B, W, and GHZ types\cite{threeM}. It was shown that these classes 
satisfy a linear hierarchy S $\subset$ B $\subset$ W $\subset$ GHZ. 

Generally, a given QIT task requires a particular type of entanglement. In addition, the effect of environment generally converts the pure state 
prepared for the QIT task into the mixed state. Therefore, it is important to distinguish the entanglement of mixtures to perform
the QIT task successfully. However, it is notoriously difficult problem to know which type of entanglement is contained in the given 
multipartite mixed state. Even for three-qubit state it is very difficult problem because analytical computation of the residual entanglement for 
arbitrary mixed states is generally impossible so far\footnote{However, it is possible to compute the residual entanglement 
for few rare cases\cite{tangle}.}.

Recently, classification of the entanglement classes for three-qubit mixed states has been significantly progressed. In Ref.\cite{elts12-1}
the GHZ symmetry was examined in three-qubit system. This is a symmetry that GHZ states 
$\ket{\mbox{GHZ}_3}_{\pm} = (1 / \sqrt{2}) (\ket{000} \pm \ket{111})$ have up to the global phase 
and is expressed as a symmetry under (i) qubit permutations, (ii) simultaneous flips, (iii) qubit rotations 
about the $z$-axis. The whole GHZ-symmetric states can be parametrized by two real parameters, {\it say} $x$ and $y$. Authors in Ref. \cite{elts12-1}
succeeded in classifying the entanglement of the GHZ-symmetric states completely. This complete classification makes it possible to 
compute the three-tangle\footnote{The definition of three-tangle in this paper is a square root of the residual entanglement presented in Ref.\cite{ckw}.} analytically 
for the whole GHZ-symmetric states\cite{siewert12-1} and to construct the class-specific optimal witnesses\cite{elts12-2}.
It also makes it possible to obtain lower bound of three-tangle for arbitrary $3$-qubit mixed state\cite{elts13-1}.  
More recently, the SLOCC classification of the extended GHZ-symmetric states was discussed\cite{jung13-1}. Extended GHZ symmetry is the GHZ 
symmetry without qubit permutation symmetry. It is larger symmetry group than usual GHZ symmetry group, and is parametrized by four real parameters.

The purpose of this paper is to extend the analysis of Ref.\cite{elts12-1} to four-qubit system. Four-qubit GHZ states\footnote{While 
$\ket{\mbox{GHZ}_3}_{+}$ is an unique maximally entangled $3$-qubit state up to  LU, $\ket{\mbox{GHZ}_4}_{+}$ is not unique maximally entangled 
state. In $4$-qubit system there are two more additional maximally entangled states $\ket{\Phi_5} = (1 / \sqrt{6}) (\sqrt{2} \ket{1111} + 
\ket{1000} + \ket{0100} + \ket{0010} + \ket{0001})$ and $\ket{\Phi_4} = (1/2) (\ket{1111} + \ket{1100} + \ket{0010} + \ket{0001}$\cite{oster06-1}.}
(or $4$-cat states\cite{bennet00}, in honor of Schr\"odinger's cat) are defined as 
\begin{equation}
\label{4q-GHZ}
\ket{\mbox{GHZ}_4}_{\pm} = \frac{1}{\sqrt{2}} (\ket{0000} \pm \ket{1111}).
\end{equation}
Like a $3$-qubit GHZ symmetry we define a $4$-qubit GHZ symmetry as a symmetry which $\ket{\mbox{GHZ}_4}_{\pm}$ have up to the global phase. 
Straightforward generalization, which is (i) qubit permutations, (ii) simultaneous flips (i.e., application of $\sigma_x \otimes \sigma_x \otimes \sigma_x \otimes \sigma_x$), (iii) qubit rotations 
about the $z$-axis of the form 
\begin{equation}
\label{four-1}
U (\phi_1, \phi_2, \phi_3) = e^{i \phi_1 \sigma_z} \otimes e^{i \phi_2 \sigma_z} \otimes e^{i \phi_3 \sigma_z}
                                \otimes e^{-i (\phi_1 + \phi_2 + \phi_3) \sigma_z},
\end{equation}
is obviously a symmetry of $\ket{\mbox{GHZ}_4}_{\pm}$. Thus, we will call this symmetry as $4$-qubit GHZ symmetry.  
As will be shown later the $4$-qubit GHZ-symmetric states are represented by 
three real parameters while $3$-qubit states contain only two. Thus, it is more difficult to analyze the entanglement 
classification in $4$-qubit GHZ-symmetric case than that in $3$-qubit case. Furthermore, if number of qubit increases,
we need real parameters more and more to represent the GHZ-symmetric states. Therefore, classification of the entanglement 
for the GHZ-symmetric states becomes a formidable task in the higher-qubit system. In this reason it is advisable to 
restrict the GHZ-symmetry to reduce the number of real parameters.
This can be achieved by modifying (ii) into 
(ii) simultaneous and any pair flips without changing (i) and (iii). In four qubit-system this modification can be stated as an 
invariance under the application of $\sigma_x \otimes \sigma_x \otimes \openone \otimes \openone$, $\sigma_x \otimes \openone \otimes \sigma_x \otimes \openone$, 
$\sigma_x \otimes \openone \otimes \openone \otimes \sigma_x$, $\openone \otimes \sigma_x \otimes \sigma_x \otimes \openone$, $\openone \otimes \sigma_x \otimes \openone \otimes \sigma_x$,
$\openone \otimes \openone \otimes \sigma_x \otimes \sigma_x$, and $\sigma_x \otimes \sigma_x \otimes \sigma_x \otimes \sigma_x$.
The simplest pure state which has the modified symmetry (ii) is 
\begin{equation}
\label{modified-2}
\ket{\psi}_{ABCD} = \frac{1}{2\sqrt{2}} (\ket{0000} + \ket{1100} + \ket{1010} + \ket{1001} + \ket{0110} + \ket{0101} + \ket{0011} + \ket{1111}).
\end{equation}
It is easy to show that $\ket{\psi}_{ABCD}$ is symmetric under the flips of $(A,B)$, $(A,C)$, $(A,D)$, $(B,C)$, $(B,D)$, $(C,D)$, or $(A,B,C,D)$ parties. 
Obviously, $\ket{\mbox{GHZ}_4}_{\pm}$ do not have this modified symmetry. Of course, the states, which have this modified symmetry, are also GHZ-symmetric. Therefore, we call this modified symmetry as 
restricted GHZ (RGHZ) symmetry\footnote{The state $\ket{\psi}_{ABCD}$ given in Eq. (\ref{modified-2}) is not RGHZ-symmetric because it is not symmetric 
under the qubit rotation about the $z$-axis although it is symmetric under the modified (ii). In fact, there is no pure RGHZ-symmetric state.}.
As will be shown, the RGHZ-symmetric states are represented by two real parameters like the $3$-qubit case.

This paper is organized as follows.  In section II the general forms of the GHZ- and RGHZ-symmetric states are derived, respectively. It is shown that
while the GHZ-symmetric states are represented by three real parameters, the RGHZ-symmetric states are represented by two real parameters. In
section III we classify the entanglement of the RGHZ-symmetric states. It is shown that entanglement of the RGHZ-symmetric states is 
either $L_{abc_2}$ or $G_{abcd}$. In section IV a brief conclusion is given.


\section{GHZ-symmetric and RGHZ-symmetric states}
In this section we will derive the general forms of the GHZ-symmetric and RGHZ-symmetric states and compare them with each other. 

\subsection{GHZ-symmetric states}
It is not difficult to show that the general form of the GHZ-symmetric states is 
\begin{eqnarray}
\label{4q-1}
& &\hspace{2.5cm} \rho_4^{\mbox{\scriptsize{GHZ}}} = \tilde{x} \left[ \ket{0000}\bra{1111} + \ket{1111} \bra{0000} \right]                            \\    \nonumber
& &+ \mbox{diag} \left(\tilde{\alpha}_1, \tilde{\alpha}_2, \tilde{\alpha}_2, \tilde{\alpha}_3, \tilde{\alpha}_2, \tilde{\alpha}_3, 
\tilde{\alpha}_3, \tilde{\alpha}_2, \tilde{\alpha}_2, \tilde{\alpha}_3, \tilde{\alpha}_3, \tilde{\alpha}_2, 
                    \tilde{\alpha}_3, \tilde{\alpha}_2, \tilde{\alpha}_2, \tilde{\alpha}_1   \right)
\end{eqnarray}
where $\tilde{x}$, $\tilde{\alpha}_1$, $\tilde{\alpha}_2$ and $\tilde{\alpha}_3$ are real numbers satisfying 
$\tilde{\alpha}_1 + 4 \tilde{\alpha}_2 + 3 \tilde{\alpha}_3 = \frac{1}{2}$. 
Unlike the three-qubit case, $\rho_4^{\mbox{\scriptsize{GHZ}}}$ is represented by three real parameters.

Now, we define following two real parameters $\tilde{y}$, $\tilde{z}$, as
\begin{eqnarray}
\label{change-1}
& &\tilde{y} = {\cal N}_1 \left[ \tilde{\alpha}_1 + (\sqrt{10} + 3) \tilde{\alpha}_2 \right]         \\   \nonumber
& &\tilde{z} = {\cal N}_2 \left[ (\sqrt{10} + 3)  \tilde{\alpha}_1 -  \tilde{\alpha}_2 \right]
\end{eqnarray}
where
\begin{equation}
\label{change-2}
{\cal N}_1 = \sqrt{\frac{2}{3} - \frac{2}{15} \sqrt{10}} \approx 0.495                  \hspace{1.0cm}
{\cal N}_2 = \sqrt{\frac{14}{3} - \frac{22}{15} \sqrt{10}} \approx 0.169.
\end{equation}
Then, it is straightforward to show that the Hilbert-Schmidt metric of $\rho_4^{\mbox{\scriptsize{GHZ}}}$ is equal to the Euclidean metric, i.e.,
\begin{equation}
\label{hilbert}
d^2 \left[ \rho_4^{\mbox{\scriptsize{GHZ}}} (\tilde{\alpha}_1, \tilde{\alpha}_2, \tilde{\alpha}_3, \tilde{x}), 
           \rho_4^{\mbox{\scriptsize{GHZ}}} (\tilde{\alpha}_1', \tilde{\alpha}_2', \tilde{\alpha}_3', \tilde{x}') \right]
= (\tilde{x} - \tilde{x}')^2 + (\tilde{y} - \tilde{y}')^2 + (\tilde{z} - \tilde{z}')^2
\end{equation}
where $d^2 (A, B) = \frac{1}{2} \mbox{tr} (A-B)^{\dagger} (A-B)$. The three real parameters $\tilde{x}$, $\tilde{y}$, and $\tilde{z}$ 
can be represented as 
\begin{eqnarray}
\label{revise1}
& &\tilde{x} = \frac{1}{2} \left[_+\bra{\mbox{GHZ}_4} \rho_4^{\mbox{\scriptsize{GHZ}}} \ket{\mbox{GHZ}_4}_+ - _-\bra{\mbox{GHZ}_4} \rho_4^{\mbox{\scriptsize{GHZ}}} \ket{\mbox{GHZ}_4}_- \right]
                                                                                                                                                                            \nonumber    \\
& &\tilde{y} = \frac{{\cal N}_1}{2} 
\bigg[\hspace{.01cm} _+\bra{\mbox{GHZ}_4} \rho_4^{\mbox{\scriptsize{GHZ}}} \ket{\mbox{GHZ}_4}_+ + _-\bra{\mbox{GHZ}_4} \rho_4^{\mbox{\scriptsize{GHZ}}} \ket{\mbox{GHZ}_4}_-               \nonumber     \\
& & \hspace{7.0cm}
+ 2 (\sqrt{10} + 3) \bra{\Phi} \rho_4^{\mbox{\scriptsize{GHZ}}} \ket{\Phi}\bigg]                                                                                                            \\   \nonumber
& &\tilde{z} = \frac{{\cal N}_2}{2} 
\bigg[\hspace{.01cm}(\sqrt{10} + 3) _+\bra{\mbox{GHZ}_4} \rho_4^{\mbox{\scriptsize{GHZ}}} \ket{\mbox{GHZ}_4}_+                              \\   \nonumber
& &  \hspace{3.0cm}
+ (\sqrt{10} + 3) _-\bra{\mbox{GHZ}_4} \rho_4^{\mbox{\scriptsize{GHZ}}} \ket{\mbox{GHZ}_4}_-   
- 2 \bra{\Phi} \rho_4^{\mbox{\scriptsize{GHZ}}} \ket{\Phi}\bigg]                                                                               
\end{eqnarray}
where $\ket{\Phi}$ is either $\ket{\Phi^+} = (\ket{0001} + \ket{1110}) / \sqrt{2}$ or $\ket{\Phi^-} = (\ket{0001} - \ket{1110}) / \sqrt{2}$.

\begin{figure}[ht!]
\begin{center}
\includegraphics[height=6cm]{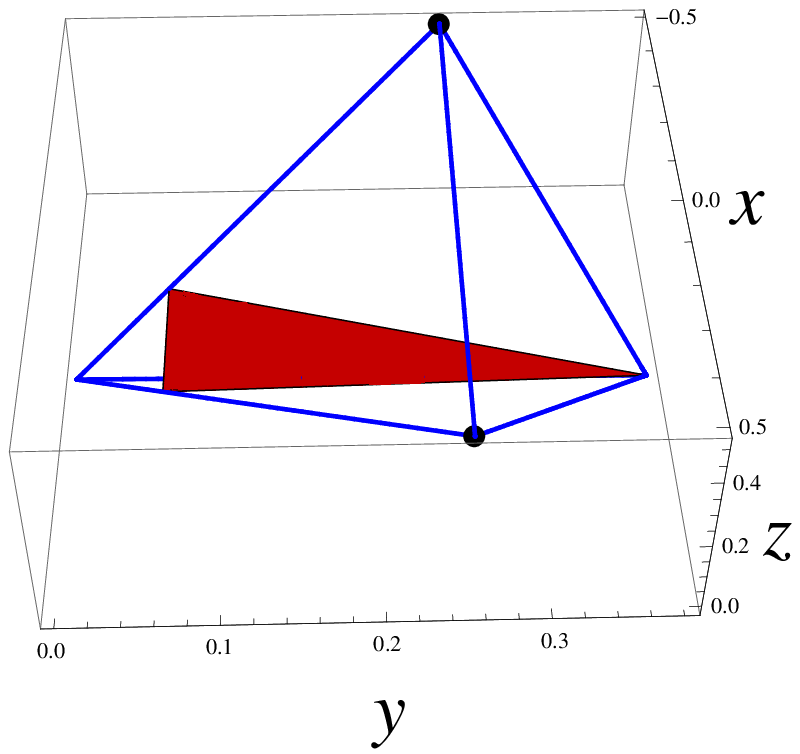}
\includegraphics[height=6cm]{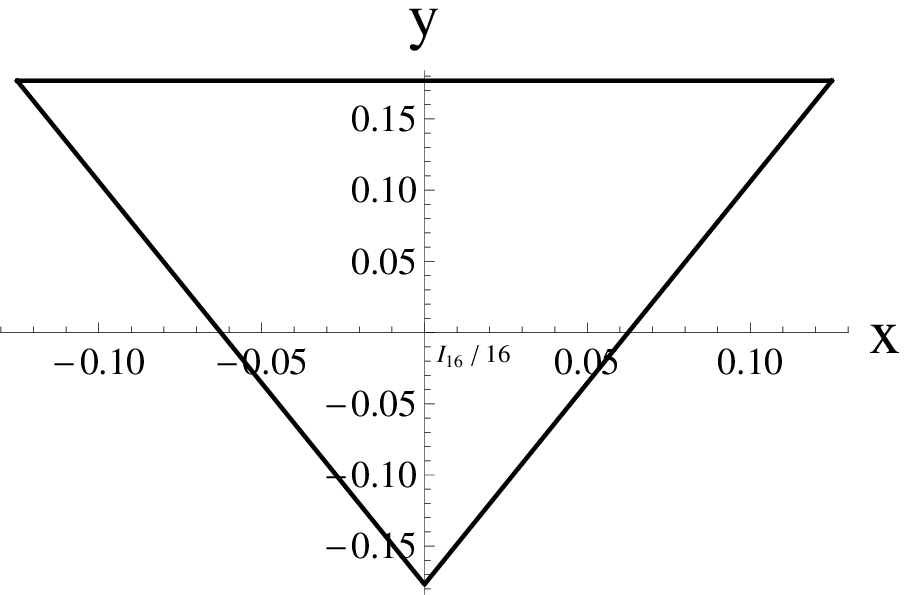}
\caption[fig1]{(Color online) Each point in tetrahedron is correspondent to the GHZ-symmetric state. Two black dots represent the $4$-qubit
GHZ state $\ket{\mbox{GHZ}_4}_{\pm}$. The surface of triangle in the tetrahedron is the place where the RGHZ-symmetric states reside. 
(b) Each point in triangle is correspondent to the RGHZ-symmetric state. This triangle is equivalent to the triangle in Fig. 1(a).
Thus, RGHZ symmetric states have very small portion and are of zero measure in the whole set of the GHZ-symmetric states.}
\end{center}
\end{figure}

In order for $\rho_4^{\mbox{\scriptsize{GHZ}}}$ to be a physical state the parameters should be restricted to
\begin{equation}
\label{restriction-1}
0 \leq \tilde{\alpha}_2 \leq \frac{1}{8}                 \hspace{1.0cm}
0 \leq \tilde{\alpha}_3 \leq \frac{1}{6}                 \hspace{1.0cm}
0 \leq \tilde{\alpha}_1 \leq \frac{1}{2}
\end{equation}
and 
\begin{equation}
\label{restriction-2}
\tilde{\alpha}_1 \geq \pm \tilde{x}.
\end{equation}
This physical conditions imply that any GHZ-symmetric physical state is represented as a point inside a tetrahedron shown in Fig. 1(a). 
In this figure two black dots represent $\ket{\mbox{GHZ}_4}_{\pm}$, respectively. It is worthwhile 
noting that the sign of $x$ does not change the character of entanglement because 
$\rho_4^{\mbox{\scriptsize{GHZ}}} (-\tilde{x}, \tilde{y}, \tilde{z}) = u \rho_4^{\mbox{\scriptsize{GHZ}}} (\tilde{x}, \tilde{y}, \tilde{z}) u^{\dagger}$, 
where $u = i \sigma_x \otimes \sigma_y \otimes \sigma_y \otimes \sigma_y$.

\subsection{RGHZ-symmetric states}
It is straightforward to show that the general form of RGHZ-symmetric states is 
\begin{eqnarray}
\label{general-form}
& &\hspace{2.5cm} \rho_4^{\mbox{\scriptsize{RGHZ}}} = x \left[ \ket{0000}\bra{1111} + \ket{1111} \bra{0000} \right]                            \\    \nonumber
& &+ \mbox{diag} \left(\alpha_1, \alpha_2, \alpha_2, \alpha_1, \alpha_2, \alpha_1, \alpha_1, \alpha_2, \alpha_2, \alpha_1, \alpha_1, \alpha_2, 
                    \alpha_1, \alpha_2, \alpha_2, \alpha_1   \right)
\end{eqnarray}
with $\alpha_1 = \frac{1}{16} + \frac{y}{2\sqrt{2}}$ and $\alpha_2 = \frac{1}{16} - \frac{y}{2\sqrt{2}}$. The parameters $x$ and $y$ are 
chosen such that the Euclidean metric in the $(x,y)$ plane coincides with the Hilbert-Schmidt metric 
$d^2 (A,B) = \frac{1}{2} \mbox{tr} (A-B)^{\dagger} (A-B)$ again. The parameters can be represented as 
\begin{eqnarray}
\label{revise11}
& &x = \frac{1}{2} \left[_+\bra{\mbox{GHZ}_4} \rho_4^{\mbox{\scriptsize{RGHZ}}} \ket{\mbox{GHZ}_4}_+ - _-\bra{\mbox{GHZ}_4} \rho_4^{\mbox{\scriptsize{RGHZ}}} \ket{\mbox{GHZ}_4}_- \right]
                                                                                                                                                                \\            \nonumber  
& &y = \sqrt{2}  
 \left[_+\bra{\mbox{GHZ}_4} \rho_4^{\mbox{\scriptsize{RGHZ}}} \ket{\mbox{GHZ}_4}_+ + _-\bra{\mbox{GHZ}_4} \rho_4^{\mbox{\scriptsize{RGHZ}}} \ket{\mbox{GHZ}_4}_-  - \frac{1}{8}\right].
\end{eqnarray}
It is also worthwhile noting that the sign of $x$ does not change the entanglement
because $\rho_4^{\mbox{\scriptsize RGHZ}} (-x, y) = u \rho_4^{\mbox{\scriptsize RGHZ}} (x,y) u^{\dagger}$. This is evident from the fact that 
the RGHZ-symmetric state is also GHZ-symmetric. 

Since $\rho_4^{\mbox{\scriptsize RGHZ}}$ is a quantum state, it 
should be a positive operator, which restricts the parameters as 
\begin{equation}
\label{restriction-3}
y \geq \pm 2 \sqrt{2} x - \frac{\sqrt{2}}{8}   \hspace{2.0cm}
|x| \leq \frac{1}{8}.
\end{equation}
Thus any RGHZ-symmetric physical state is represented as a point in a triangle depicted in Fig. 1(b). 

It is easy to show that $\rho_4^{\mbox{\scriptsize{GHZ}}}$ is RGHZ-symmetric if and only if $\tilde{x} = x$, $\tilde{\alpha}_2 = \alpha_2$, and 
$\tilde{\alpha_1} = \tilde{\alpha}_3 = \alpha_1$ or equivalently
\begin{equation}
\label{compare-1}
\tilde{y} = {\cal N}_1 \left[ \frac{\sqrt{10} + 4}{16} - \frac{\sqrt{10} + 2}{2\sqrt{2}} y \right]    
\hspace{1cm}
\tilde{z} = {\cal N}_2 \left[\frac{\sqrt{10} + 2}{16} + \frac{\sqrt{10} + 4}{2\sqrt{2}} y \right].
\end{equation}
Using this relation it is possible to know where the RGHZ-symmetric states reside in the tetrahedron in Fig. 1(a). In this figure the red 
triangle is equivalent one of Fig. 1(b). Thus, the states on this triangle are RGHZ-symmetric. From Fig. 1(a)
one can realize that the RGHZ-symmetric states have very small portion and are of zero measure in the entire set of GHZ-symmetric states.

\section{SLOCC Classification of RGHZ-symmetric States}
The SLOCC classification of the $4$-qubit pure-state system was first discussed in \cite{fourP-1} by making use of the Jordan block structure
of some complex symmetric matrix. Subsequently, same issue was explored in several more papers using different approaches\cite{fourP-2}.
Unlike, however, two- and three-qubit cases, the results of Ref.\cite{fourP-1,fourP-2} seem to be contradictory to each other.
This means that still our understanding on the $4$-qubit entanglement is incomplete.
 
In this paper we adopt the results of \cite{fourP-1}, where there are following nine inequivalent SLOCC classes;
\begin{eqnarray}
\label{slocc-ver}
G_{abcd}&=&\frac{a+d}{2}(|0000\rangle
+|1111\rangle)+\frac{a-d}{2}(|0011\rangle
+|1100\rangle)                                                  \nonumber  \\   
&&\hspace{.1cm}+\frac{b+c}{2}(|0101\rangle
+|1010\rangle)+\frac{b-c}{2}(|0110\rangle +|1001\rangle)                      \nonumber   \\  
L_{abc_2}&=&\frac{a+b}{2}(|0000\rangle
+|1111\rangle)+\frac{a-b}{2}(|0011\rangle +|1100\rangle)                             \nonumber   \\ 
&&\hspace{.5cm}+c(|0101\rangle +|1010\rangle)+|0110\rangle                            \nonumber   \\   
L_{a_2b_2}&=&a(|0000\rangle +|1111\rangle)+b(|0101\rangle+|1010\rangle)             
+|0110\rangle +|0011\rangle                                             \nonumber   \\
L_{ab_3}&=&a(|0000\rangle
+|1111\rangle)+\frac{a+b}{2}(|0101\rangle
+|1010\rangle)                                                                              \\
&&\hspace{.1cm}+\frac{a-b}{2}(|0110\rangle +|1001\rangle)                                   
+\frac{i}{\sqrt{2}}(|0001\rangle +|0010\rangle
+|0111\rangle
+|1011\rangle)                                                                               \nonumber    \\
L_{a_4}&=&a(|0000\rangle +|0101\rangle +|1010\rangle
+|1111\rangle)                                                                           
+(i|0001\rangle +|0110\rangle -i|1011\rangle)                            \nonumber    \\
L_{a_20_{3\oplus\bar{1}}}&=&a(|0000\rangle
+|1111\rangle)+(|0011\rangle +|0101\rangle +|0110\rangle)                                 \nonumber   \\
L_{0_{5\oplus\bar{3}}}&=&|0000\rangle +|0101\rangle +|1000\rangle
+|1110\rangle                                                                          \nonumber          \\
L_{0_{7\oplus\bar{1}}}&=&|0000\rangle +|1011\rangle +|1101\rangle
+|1110\rangle                                                                            \nonumber        \\
L_{0_{3\oplus\bar{1}}0_{3\oplus\bar{1}}}&=&|0000\rangle+|0111\rangle,                      \nonumber
\end{eqnarray}    
where $a$, $b$, $c$, and $d$ are complex parameters with nonnegative real part.
In Eq. (\ref{slocc-ver}) $G_{abcd}$ is special in a sense that its all local states are completely mixed. In other words, $G_{abcd}$ is 
a set of normal states\cite{verst03}.

\subsection{$L_{abc_2}$}
\begin{figure}[ht!]
\begin{center}
\includegraphics[height=10cm]{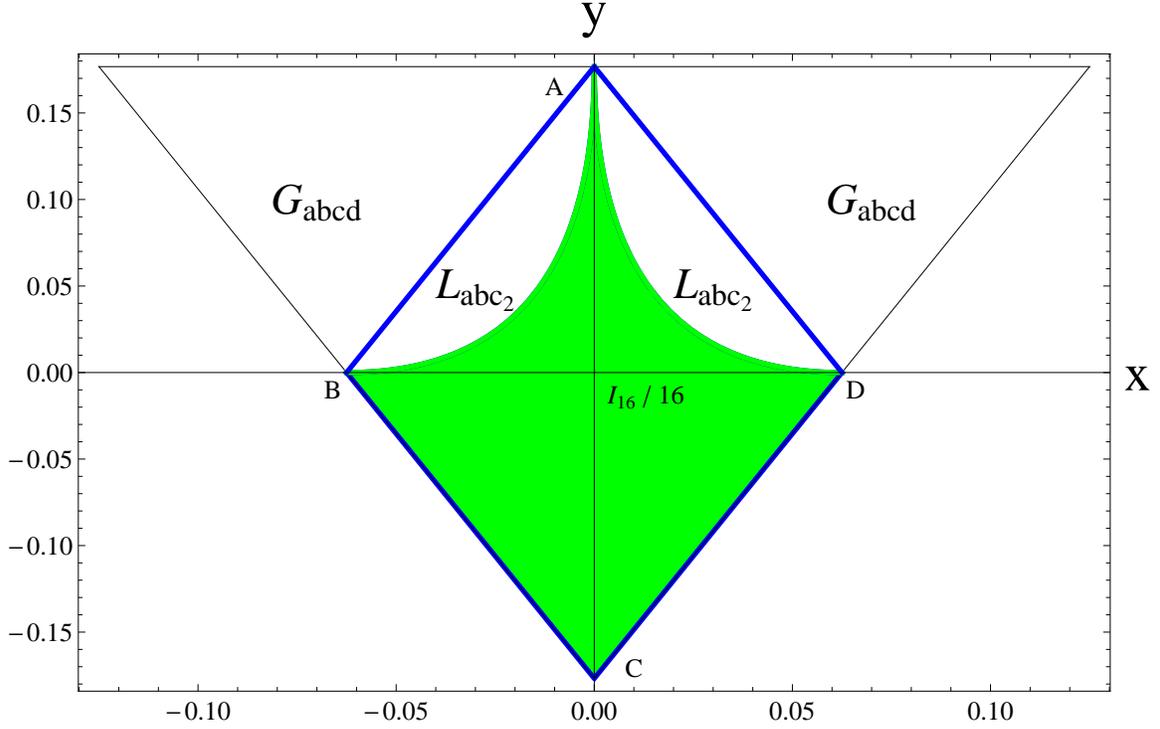}
\caption[fig1]{(Color online) The SLOCC classification of RGHZ-symmetric states $\rho_4^{\mbox{\scriptsize{RGHZ}}}$. In the polygon ABCD states of 
$L_{abc_2}$ reside. Theorem 2 implies that there is no one-qubit tensor product three-qubit entangled states in the GHZL-symmetric
states. This fact implies that the RGHZ symmetry exclude $L_{a_20_{3\oplus\bar{1}}}$, $L_{0_{3\oplus\bar{1}}0_{3\oplus\bar{1}}}$, and 
$L_{a_2b_2}$. Theorem 3 implies that there are states of $G_{abcd}$ outside the polygon ABCD.}
\end{center}
\end{figure}

In this subsection we examine a question where the states of $L_{abc_2}$ reside in the triangle in Fig. 1(b). 
Before proceeding further, it is important to note that there is a correspondence between four-qubit pure states and RGHZ-symmetric states. Let $\ket{\psi}$ be a 
four-qubit pure state. Then, the corresponding RGHZ-symmetric state $\rho_4^{\mbox{\scriptsize{RGHZ}}} (\psi)$ can be written as 
\begin{equation}
\label{4q-2}
\rho_4^{\mbox{\scriptsize{RGHZ}}} (\psi) = \int U \ket{\psi} \bra{\psi} U^{\dagger},
\end{equation}
where the integral is understood to cover the entire RGHZ symmetry group, i.e., unitaries $U (\phi_1, \phi_2, \phi_3)$ in Eq. (\ref{four-1}) and 
averaging over the discrete symmetries. For example, if $\ket{\psi} = \sum_{i,j,k,l = 0}^1 \psi_{ijkl} \ket{ijkl}$, 
$\rho_4^{\mbox{\scriptsize{RGHZ}}} (\psi)$ becomes Eq. (\ref{general-form}) with
\begin{eqnarray}
\label{4q-3}
& &x = \frac{1}{4} \mbox{Re} \bigg[ \psi_{0000} \psi_{1111}^* + \psi_{0011} \psi_{1100}^* + \psi_{0101} \psi_{1010}^* + \psi_{0110} \psi_{1001}^* \bigg]
                                                                                                                                \\   \nonumber
& &\alpha_1 \equiv \frac{1}{16} + \frac{y}{2\sqrt{2}} = 
\frac{1}{8} \bigg[|\psi_{0000}|^2 + |\psi_{1111}|^2 + |\psi_{0011}|^2 + |\psi_{0101}|^2                                          \\    \nonumber
& & \hspace{4.0cm}
+ |\psi_{0110}|^2 + |\psi_{1001}|^2 + |\psi_{1010}|^2 + |\psi_{1100}|^2 \bigg].
\end{eqnarray}

Now, we are ready to discuss the main issue of this subsection.

{\bf Theorem 1.} {\it The RGHZ-symmetric states of $L_{abc_2}$-class reside in the polygon ABCD in Fig. 2.} 

\smallskip

{\bf Proof.}
First we note that when $a=b=c=0$, $L_{abc_2}$ reduces to the fully separable state $\ket{0110}$. 
Since LU is a particular case of SLOCC, this fact implies that all fully separable states are in the $L_{abc_2}$.
Let $\ket{\psi^{sep}} = (u_1 \otimes u_2 \otimes u_3 \otimes u_4) \ket{0000}$, where
\begin{eqnarray}
\label{4sep-1}
u_j = \left(          \begin{array}{cc}
                 A_j  &  -C_j^*      \\
                 C_j  &  A_j^*
                       \end{array}           \right)     \hspace{1.0cm} \mbox{with} \hspace{.2cm} |A_j|^2 + |C_j|^2 = 1.
\end{eqnarray}
Then, it is easy to derive the parameters $x$ and $y$ of $\rho_4^{\mbox{\scriptsize{RGHZ}}} (\psi^{sep})$ easily using Eq. (\ref{4q-3}). 
Our method for proof is as follows. Applying the Lagrange multiplier method we maximize $x$ with 
given $y$. Then, it is possible to derive a boundary $x_{\max} = x_{\max} (y)$ in the $(x,y)$ plane. 
If a region inside the boundary is convex, this is the region where the 
$L_{abc_2}$-class states reside. If it is not convex, we have to choose the convex hull of it for the residential region. 

From a symmetry it is evident that the maximum of $x$ occurs when $A_1 = A_2 = A_3 = A_4 \equiv A$. Then the constraint of $y$ yields
$A^2 = \frac{1}{2} \left(1 \pm 2^{5/8} y^{1/4} \right)$, which gives
\begin{equation}
\label{4sep-2}
x_{max} = \frac{1}{16} \left(1 - 2^{\frac{5}{4}} y^{\frac{1}{2}} \right)^2.
\end{equation}
Since the sign of $x_{max}$ does not change the entanglement class, the region represented by green color in Fig. 2 is derived. Since it is 
not convex, we have to choose a convex hull, which is a polygon ABCD in Fig. 2. This completes the proof.
 
Although we start with fully separable state, this does not guarantee that all states in the polygon ABCD are fully separable because 
$L_{abc_2}$ has $4$-way entangled states as well as fully separable states. The only fact we can assert is that all $L_{abc_2}$-class
RGHZ-symmetric states reside in the polygon ABCD. 

\subsection{$L_{a_20_{3\oplus\bar{1}}}, L_{0_{3\oplus\bar{1}}0_{3\oplus\bar{1}}}, \cdots $}
In this subsection we will show that the RGHZ symmetry excludes all SLOCC classes except $G_{abcd}$.  

{\bf Theorem 2.} {\it There is no one-qubit product GHZ state in the RGHZ-symmetric states.} 

\smallskip

{\bf Proof.}
Let $\ket{\psi^{GHZ}} = (G_1 \otimes G_2 \otimes G_3 \otimes G_4) \ket{0} \otimes (\ket{000} + \ket{111})$, where
\begin{eqnarray}
\label{4ghz-1}
G_j = \left(        \begin{array}{cc}
                     A_j  &  B_j       \\
                     C_j  &  D_j
                     \end{array}              \right).
\end{eqnarray}
Then, it is easy to compute $x$ and $y$ of $\rho_4^{\mbox{\scriptsize{RGHZ}}} (\psi^{GHZ})$ using Eq. (\ref{4q-3}). Now, we want to 
maximize $x$ with given $y$ and $\bra{\psi^{GHZ}} \psi^{GHZ} \rangle = 1$. From symmetry of the Lagrange multiplier equations 
it is evident that the maximum of $x$
occurs when $A_2 = A_3 = A_4 = B_2 = B_3 = B_4 \equiv a$ and $C_2 = C_3 = C_4 = D_2 = D_3 = D_4 \equiv c$. Then, we define
$x^{\Lambda} = x + \Lambda_0 \Theta_0 + \Lambda_1 \Theta_1$, where $\Lambda_0$ and $\Lambda_1$  are Lagrange multiplier constants, and
\begin{eqnarray}
\label{4ghz-2}
& &x = 4 A_1 C_1 a^3 c^3                                                     \\    \nonumber
& &\Theta_0 = 4 (A_1^2 + C_1^2) (a^2 + c^2)^2 - 1                     \\    \nonumber
& &\Theta_1 = 4 \left[ A_1^2 a^2 (a^4 + 3 c^4) + C_1^2 c^2 (3a^4 + c^4) - 2 \alpha_1 \right].
\end{eqnarray}
Now, we want to maximize $x$ under the constraints $\Theta_0 = \Theta_1 = 0$. 

First, we solve the two constraints, whose solutions are 
\begin{equation}
\label{4ghz-3}
A_1^2 = \frac{8 \alpha_1 (u_1 + u_2) - u_2}{u_1^2 - u_2^2}                       \hspace{2.0cm}
C_1^2 = \frac{u_1 - 8 \alpha_1 (u_1 + u_2)}{u_1^2 - u_2^2},
\end{equation}
where $u_1 = 4 a^2 (a^4 + 3 c^4)$ and $u_2 = 4 c^2 (3 a^4 + c^4)$. 
From $\frac{\partial x^{\Lambda}}{\partial A_1} = \frac{\partial x^{\Lambda}}{\partial C_1} = 0$ one can express the Lagrange multiplier 
constants as 
\begin{equation}
\label{4ghz-4}
\Lambda_0 = - \frac{A_1^2 u_1 - C_1^2 u_2}{A_1 C_1} \frac{2 a^3 c^3}{u_1^2 - u_2^2}      \hspace{2.0cm}
\Lambda_1 = \frac{A_1^2 - C_1^2}{A_1 C_1} \frac{2 a^3 c^3}{u_1 - u_2}.
\end{equation}
Combining Eqs. (\ref{4ghz-3}), (\ref{4ghz-4}), and $\frac{\partial x^{\Lambda}}{\partial a} = \frac{\partial x^{\Lambda}}{\partial c} = 0$,
we obtain 
\begin{equation}
\label{4ghz-5}
8 \alpha_1 (z^2 + 1)^4 = z^8 + 6 z^4 + 1,
\end{equation}
where $z = \frac{a}{c}$. Then, the maximum of $x$ with given $y$ becomes
\begin{equation}
\label{4ghz-6}
x_{max} = \frac{z^3 \sqrt{8 \alpha_1 (1 - 8 \alpha_1) (1 + z^2)^6 - z^2 (3 + z^4) (1 + 3 z^4)}}{(z^4 - 1)^3}.
\end{equation}

Using Eq. (\ref{4ghz-5}) and performing long and tedious calculation, one can show that the right-hand side of Eq. (\ref{4ghz-6}) reduces to 
$\frac{1}{16} \left(1 - \sqrt{16 \alpha_1 - 1} \right)^2$, which results in the identical equation with Eq. (\ref{4sep-2}). This implies that 
there is no one-qubit product three-qubit GHZ state in the RGHZ-symmetric states. This completes the proof.

From this theorem one can conclude that there is no $L_{0_{3\oplus\bar{1}}0_{3\oplus\bar{1}}}$ in the RGHZ-symmetric states, because 
this class involves one-qubit product GHZ-state. Since it is well-known that the three-qubit states consist of fully separable (S), 
bi-separable (B), W, and GHZ states, and they satisfy
a linear hierarchy  S $\subset$ B $\subset$ W $\subset$ GHZ, theorem 2 also implies that there is no one-qubit product W state
in the RGHZ-symmetric states. Thus, RGHZ symmetry excludes $L_{a_20_{3\oplus\bar{1}}}$ too because this class contains one-qubit product W state
when $a=0$. This theorem also implies that there is no one-qubit product one-qubit product B state, which excludes $L_{a_2b_2}$.
Similarly, one can exclude all classes except $G_{abcd}$-class\footnote{For other classes it is more easy to adopt the following numerical calculation
than applying the Lagrange multiplier method. First, we select a representative state $\ket{\psi}$ for each SLOCC class. Next, we generate $16$ random 
numbers and identify them with $A_j, B_j, C_j, D_j \hspace{.3cm} (j = 1, \cdots, 4)$. Then, using a mapping (\ref{4q-3}) one can compute 
$x$ and $y$ for pure state $G_1 \otimes G_2 \otimes G_3 \otimes G_4 \ket{\psi}$. Repeating this procedure over and over and collecting all
$(x, y)$ data, one can deduce numerically the residential region of this class. The numerical calculation shows that the residence of all SLOCC class except $G_{abcd}$ is confined in the polygon ABCD of Fig. 2.}.

\subsection{$G_{abcd}$}
Now, we want to discuss the entanglement classes of remaining RGHZ-symmetric states. In order to conjecture the classes quickly, let 
us consider the following double bi-separable state
\begin{equation}
\label{bb-1}
\ket{\psi^{BB}} = \frac{1}{\sqrt{2}} \left(\ket{00} + \ket{11} \right) \otimes \frac{1}{\sqrt{2}} \left(\ket{00} + \ket{11} \right).
\end{equation}
Such a state belongs to $G_{abcd}$ with $(a=1, b=c=d=0)$ or $a=b=c=d$.  
Then, Eq. (\ref{4q-3}) shows that the parameters of $\rho_4^{\mbox{\scriptsize{GHZL}}} (\psi^{BB})$ are $x = 1/8$ and $y = \sqrt{2} / 8$, 
which correspond to the right-upper corner of the triangle in Fig. 2. Since mixing can result only in the same or a lower entanglement
class, the entanglement class of this corner state should be $G_{abcd}$ or its sub-classes. However, the sub-class of this state should be a class, 
where fully separable states belong, and
those states are confined in $ABCD$. Therefore, the corner should be $G_{abcd}$. This fact strongly suggests that all remaining states in Fig. 2 
are $G_{abcd}$. The following theorem shows that our conjecture is correct.

{\bf Theorem 3.} {\it All remaining RGHZ-symmetric states in Fig. 2 are $G_{abcd}$-class.} 

\smallskip

{\bf Proof.}
Let $\ket{\psi^{BB}} = (G_1 \otimes G_2 \otimes G_3 \otimes G_4) (\ket{00} + \ket{11}) \otimes (\ket{00} + \ket{11})$, where
$G_j$ is given in Eq. (\ref{4ghz-1}). Then, it is easy to compute the parameters $x$ and $y$ of $\rho_4^{\mbox{\scriptsize{GHZL}}} (\psi^{BB})$
using Eq. (\ref{4q-3}). Similar to the previous theorems we want to maximize $x$ with given $y$. From a symmetry of Lagrange multiplier equations
it is evident that the 
maximum of $x$ occurs when
\begin{eqnarray}
\label{bb-3}
& &A_1 = A_2 \equiv a_1   \hspace{1.5cm}  A_3 = A_4 \equiv a_3                          \\    \nonumber
& &B_1 = B_2 \equiv b_1   \hspace{1.5cm}  B_3 = B_4 \equiv b_3                          \\    \nonumber
& &C_1 = C_2 \equiv c_1   \hspace{1.5cm}  C_3 = C_4 \equiv c_3                          \\    \nonumber
& &D_1 = D_2 \equiv d_1   \hspace{1.5cm}  D_3 = D_4 \equiv d_3.
\end{eqnarray}                         
For later convenience we define $\mu_1 = a_1^2 + b_1^2$, $\mu_2 = a_3^2 + b_3^2$, $\mu_3 = c_1^2 + d_1^2$, $\mu_4 = c_3^2 + d_3^2$, 
$\nu_1 = a_1 c_1 + b_1 d_1$, and $\nu_2 = a_3 c_3 + b_3 d_3$. 

In order to apply the Lagrange multiplier method we define $x^{\Lambda} = x + \Lambda_0 \Theta_0 + \Lambda_1 \Theta_1$, where
\begin{eqnarray}
\label{bb-4}
& &x = \frac{1}{2} \left(\mu_1 \mu_2 \mu_3 \mu_4 + \nu_1^2 \nu_2^2 \right)               \\   \nonumber
& &\Theta_0 = (\mu_1^2 + 2 \nu_1^2 + \mu_3^2) (\mu_2^2 + 2 \nu_2^2 + \mu_4^2) - 1        \\    \nonumber
& &\Theta_1 = (\mu_1^2 + \mu_3^2) (\mu_2^2 + \mu_4^2) + 4 \nu_1^2 \nu_2^2 - 8 \alpha_1.
\end{eqnarray}
The constraints $\Theta_0 = 0$ and $\Theta_1 = 0$ come from $\bra{\psi^{BB}} \psi^{BB} \rangle = 1$ and Eq. (\ref{4q-3}), respectively.

Now, we have eight equations $\frac{\partial x^{\Lambda}}{\partial \mu_i} = 0 \hspace{.2cm} (i = 1, 2, 3, 4)$, 
$\frac{\partial x^{\Lambda}}{\partial \nu_i} = 0 \hspace{.2cm} (i = 1, 2)$, and $\Theta_0 = \Theta_1 = 0$. Analyzing those equations, one can show 
that the maximum of $x$ occurs when $\mu_1 = \mu_3$ and $\mu_2 = \mu_4$. Then, the constraint $\Theta_1 = 0$ implies 
\begin{equation}
\label{bb-5}
x_{max} = \frac{1}{16} + \frac{y}{2 \sqrt{2}},
\end{equation}
which corresponds to the right side of the triangle in Fig. 2. This fact implies that the whole RGHZ-symmetric states are $G_{abcd}$ or its 
sub-class. Since $L_{abc_2}$ are confined in the polygon ABCD and the remaining classes except $G_{abcd}$ are already excluded, 
the states outside the polygon ABCD should be $G_{abcd}$-class, which completes the proof.

Although we start with a double bi-separable state, this fact does not implies that all states outside the polygon are double bi-separable
because $G_{abcd}$ contains $4$-way entangled states as well as double bi-separable states. The only fact we can say is that all states 
outside the polygon ABCD are $G_{abcd}$-class.

\section{conclusion}

In this paper the GHZ and RGHZ symmetries in four-qubit system are examined. It is shown that the whole RGHZ-symmetric states involve only 
two SLOCC classes, $L_{abc_2}$ and $G_{abcd}$. 
Following Ref. \cite{elts12-2} we can use our result to construct the optimal witness $\mathcal{W}_{\mathrm{G_{abcd}\setminus L_{abc_2}}}$, which can 
detect the $G_{abcd}$-class optimally from a set of $L_{abc_2}$ plus $G_{abcd}$ states. 

As remarked earlier if we choose GHZ symmetry, the symmetric states are represented by 
three real parameters as Eq. (\ref{4q-1}) shows. Probably, these symmetric states involve more kinds of the four-qubit SLOCC classes.
The SLOCC classification of Eq. (\ref{4q-1}) will be explored in the future. 

 Another interesting extension of present paper is to generalize our analysis to any $2n$-qubit system. Then, our modification of 
 symmetry should be changed into `any one-pair, two-pair, $\cdots$, and $n$-pair flips'. This would drastically
 reduce the number of free parameters in the set of symmetric states. 
 This strongly restricted symmetry may shed light on the SLOCC classification of the multipartite 
 states.


{\bf Acknowledgement}:
This work was supported by the Kyungnam University Foundation Grant, 2013.

\end{document}